\documentclass{aa}
\usepackage{graphicx}
\usepackage{amsmath}   % For getting proper math eqns
\usepackage{amssymb}
\usepackage{times}

\def\eq#1{{Eq.~(\ref{#1})}}

\begin{document}

\title{Contribution of Flares from Tidal Disruption of Stars to high-redshift AGN}
\titlerunning{}

 \author{Hamsa Padmanabhan
          \inst{1}
          \and
          Abraham Loeb\inst{2}
          }

   \institute{D\'epartement de Physique Th\'eorique, Universit\'e de Gen\`eve \\
24 quai Ernest-Ansermet, CH 1211 Gen\`eve 4, Switzerland\\
              \email{hamsa.padmanabhan@unige.ch}
         \and
            Astronomy department, Harvard University \\
60 Garden Street, Cambridge, MA 02138, USA \\
             \email{aloeb@cfa.harvard.edu}            
             }

   \date{}

\abstract
{We explore the contributions of Tidal Disruption Events (TDEs)  {{to the flares}} in Active Galactic Nuclei (AGN) { at high redshifts}.  Using the latest data available from X-ray and optical observations of high-redshift galaxies, in combination with the evolution of their central supermassive black holes, we calculate the contribution of TDE  to AGNs as a function of their luminosities. We find that at low redshifts ($z < 1$), a few percent of all AGN with bolometric luminosities $L_{\rm Bol} \lesssim 10^{44}$ erg s$^{-1}$ may be attributable to possible TDEs. However, this fraction can significantly increase at earlier cosmic times, including up to several tens of percent of the {population of AGN} at $z \gtrsim 3$. TDEs may comprise a significant fraction of the Compton-Thick (CT) AGN population at $z \gtrsim 3$. The above findings motivate further calibration with upcoming X-ray missions and spectroscopic surveys targeting TDE-AGN.}


\maketitle

\section{Introduction}
Supermassive black holes (SMBHs) are known to exist at the centres of galaxies \citep[for reviews, see, e.g.,][]{kormendy2013, graham2016}.  The luminosity function of Active Galactic Nuclei (AGN) constrains the growth of SMBHs that accrete at high efficiencies for extended periods of time. However, the majority of galactic nuclei are dormant, and low luminosity AGN are difficult to study due to incomplete knowledge of their accretion processes. One of the major unsolved problems in AGN  research involves the fuelling of low-luminosity AGN \citep[e.g.,][]{martini2004} and the growth history of SMBHs in the early universe.  

A key contribution to the fuelling of AGN, especially at low luminosities, could be due to Tidal Disruption Events (TDEs).
These occur when stars are deflected to a distance where they are shredded by the tidal force of the supermassive black hole when it exceeds their self-gravity \citep{rees1988, hills1975}. 
A significant fraction (roughly one-half) of the tidally disrupted stellar material is subsequently accreted on to the central black hole, producing a luminous flare at typically super-Eddington efficiencies \citep[e.g.,][]{loeb1997,kara2016}. The characteristic temporal signature of a TDE is the fall-off of the initial flare with time $t$ over a period of months to years \citep[e.g.,][]{strubbe2009} as $t^{-5/3}$. However, if the mass of the central black hole exceeds\footnote{{A fully general relativistic calculation finds that the  maximum mass required for tidal disruption to occur \citep{hills1975}} is about a few times $10^8 M_{\odot}$ for a spinning black hole, compared to about a few times $10^7 M_{\odot}$ obtained by analytical calculations for non-spinning black holes \citep[e.g.,][]{rossi2020}.} about $10^8 M_{\odot}$, a solar-type star is swallowed whole \citep{kesden2012}, with no radiation being produced.

So far, there have been approximately 83 observed flares that {may be} consistent with {{possible}} TDEs\footnote{A live catalog of observed TDE events is provided in https://tde.space/statistics/host-galaxies/}, spanning the radio, optical, UV, hard X-ray and soft X-ray bands \citep[e.g.,][]{bloom2011, levan2011, cenko2012, vanvelzen2011, gezari2012, alexander2020, vanvelzen2021}. The number is expected to increase by orders of magnitude with current and upcoming surveys, e.g. the Panoramic Survey Telescope and Rapid Response System (PanSTARRS\footnote{http://pan-starrs.ifa.hawaii.edu/public/}), the Zwicky Transient Factory (ZTF\footnote{https://www.ztf.caltech.edu/}) and the Rubin Observatory Legacy Survey of Space and Time (LSST\footnote{http://www.lsst.org}). Detailed simulations have been used to evaluate the dominant physical processes governing TDE flares in different host galaxies \citep[e.g.,][]{stone2011, stone2012, stone2016, guillochon2015}. It is believed that TDEs likely contributed to the fuelling and growth of Intermediate Mass Black Holes (IMBHs), which are considered the `missing link' between stellar mass black holes and SMBHs \citep[e.g.,][]{Sakurai2019, fialkov2017, alexander2017} { and may have recently been discovered as the remnant of the GW190521 merger \citep[150 $M_{\odot}$;][]{ligo2020, ligo2020a}}. Stars swallowed whole still contributed to the growth of SMBHs above $10^8 M_{\odot}$.  {{TDEs could make a significant contribution to the observed Changing-Look (CL) phenomenon in AGN, as noted by \citet{merloni2015}, and also explain recent trends seen in Compton-Thick (CT) AGN \citep[][]{lanzuisi2018}.}}  Observations of TDEs at high redshifts, therefore, provide valuable clues towards understanding the origin of the first black holes in the universe.

In this Letter, we use an empirical approach for constraining the contribution of TDEs to fuelling AGN at early times. We extend the locally inferred relations between {{the central black hole mass and stellar mass of the host galaxy to high redshifts \citep{kormendy2013, nobuta2012, caplar2015}}} to study the fraction of TDEs that may be expected to {contribute to} AGN of various bolometric luminosities as a function of redshift. We explore the dependence of these results on specific AGN populations, such as Changing-Look Quasars \citep[CLQs; e.g.][]{macleod2019, yang2018, stern2018, ross2018} and Compton Thick \citep[CT; e.g.][]{malizia2009, Balokovi2014, marchesi2018} AGN. Our results provide a useful benchmark in combining the high-redshift TDE data anticipated from upcoming time domain optical surveys (such as LSST) with those from X-ray studies of AGN populations. Further, these findings can also be used to place important constraints on the expected gravitational wave event rates associated with binary IMBH and SMBH mergers, detectable by the forthcoming \textit{Laser Interferometer Space Antenna} (LISA\footnote{http://lisa.nasa.gov/}) mission.

This paper is organized as follows. In Sec. \ref{sec:formalism}, we review the basic formalism connecting TDE rates to the properties of their host galaxies and central black holes. We describe the probability of TDEs arising in AGN as a function of their bolometric luminosity, accretion efficiency and host galaxy properties, and extend this framework to high redshifts. In Sec. \ref{sec:data}, we combine the TDE probabilities with the latest compilations of absorbed and unabsorbed AGN luminosity functions \citep{aird2015} and the empirical evolution of the stellar masses in galaxies. We 
apply this to specific cases of AGN populations in Sec. \ref{sec:populations} to estimate the contribution of TDE  to Changing Look Quasars (CLQs) and Compton Thick (CT) AGN. Finally, we summarize our findings and discuss future observational prospects in Sec. \ref{sec:conclusions}.

\section{TDEs at high redshift}
\label{sec:formalism}
{{We start with the analytical expression for the distribution function of TDEs in AGN, defined as the the probability of a galaxy hosting a black hole of mass $M_{\rm BH}$ whose bolometric luminosity lies between {$\log_{10} L_{\rm bol}$ and $\log_{10} L_{\rm bol} + d \log_{10} L_{\rm bol}$}, activated due to a TDE event. This probability is given by}}
\citep[e.g.,][]{merloni2015}, 
\begin{equation}
 p_{\rm TDE} (L_{\rm bol}) = \frac{\Gamma_{\rm TDE} t_{\rm peak}}{\gamma_{\rm TDE}} \exp(-L_{\rm 
bol}/L_{\rm peak}) \left(\frac{L_{\rm bol}}{L_{\rm peak}}\right)^{-1/\gamma_{\rm 
TDE}} \, ,
 \label{ptdestart}
\end{equation}
{ The above equation is derived by assuming that the AGN light curve is composed of two main phases: a `peak’ phase, where the luminosity rises rapidly and saturates at a peak value for a time $t_{\rm peak}$, and a ‘declining’ phase, where the light curve follows a power-law decline with the exponent $\gamma_{\rm TDE} = 5/3$. \footnote{{We use the standard power-law decline here \citep[e.g.,][]{evans1989}, though we note that this can be a function of time and stellar structure \citep[e.g.,][]{auchettl2017}.}} In place of the generic 
function $f(L_{\rm bol}/L_{\rm peak})$ assumed by \citet{merloni2015} to characterize the declining phase, we have used 
$f(L_{\rm bol}/L_{\rm peak}) = \exp(-L_{\rm bol}/L_{\rm peak})$ (the function 
$f$ is assumed to asymptote to near-unity for $L_{\rm bol} \ll L_{\rm peak}$ and 
drops exponentially at large $L_{\rm bol}$, so the exponential form represents a natural 
choice for the purposes of the present work.)
The term $\Gamma_{\rm TDE}$ is the rate of TDEs with peak luminosity $L_{\rm peak}$ 
occurring over a time scale $t_{\rm peak}$ in an AGN of bolometric luminosity 
$L_{\rm bol}$.}
{ {For an individual galaxy, the bolometric luminosity, $L_{\rm bol}$, may be related to the mass of 
the central supermassive black hole, $M_{\rm BH}$, by the relation:
\begin{eqnarray}
L_{\rm Bol} &=& 1.38 \times 10^{38} \eta \left(\frac{M_{\rm BH}}{M_{\odot}}\right) \  {\mathrm{erg \ s}}^{-1} \nonumber \\
&=& \eta L_{\rm Edd} \, ,
\end{eqnarray}
where we have introduced the Eddington ratio $\eta$ and the Eddington luminosity $L_{\rm Edd}$ of the central black hole. For a population of active galaxies, the distribution of the Eddington ratios of the central supermassive black hole may be modelled using Gaussian/log-normal or Schechter functional forms \citep[e.g.,][]{weigel2017, kelly2013, schulze2015}. A least-squares fit to observations of the Eddington Ratio Distribution Function (ERDF) of Broad-Line AGN from the SXDS survey over $z \sim 1.4$ \citep{nobuta2012} {{favours a lognormal form with the following mean Eddington ratio as a  function of $L_{\rm Bol}$}}:
\begin{equation}
 \overline{\log {\eta}} = 0.469 \log (L_{\rm Bol}/\mathrm{erg \ s^{-1}}) - 22.46
\end{equation}
where $\overline{\log \eta}$ is the mean of the  Eddington ratio (expressed in log units) of the sample}}. From this, the relation between the bolometric luminosity and the black hole mass can be expressed as:
\begin{equation}
 \log (L_{\rm Bol}/\mathrm{erg \ s^{-1}}) = 29.53 + 1.88 \log (M_{\rm BH}/M_{\odot})
 \label{lbolmbh}
\end{equation}
The scatter in the above relation is found to be 0.4 dex, which reflects the distribution of the {{FWHM of the lognormal distribution \citep{nobuta2012}}}.

We can generalize the above expression to high redshifts using the relation 
between the black hole mass and the host galaxy stellar
mass, $M_*$ of the galaxy\footnote{Note that we follow \citet{kormendy2013} in ignoring the distinction between the bulge and total stellar mass when considering the evolution in a statistical sense.}, which is constrained from local observations \citep[][see Sec. 6.10 and Eq. 10 of that paper]{kormendy2013}  to have the form:
\begin{eqnarray}
\frac{M_{\rm BH}}{10^9 M_{\odot}} =   \left(0.49 ^{+0.06}_{-0.05}\right) \left(\frac{M_*}{10^{11} M_{\odot}}\right)^{1.16 \pm 0.08}
\label{mbhmstar}
\end{eqnarray}
with an intrinsic scatter of 0.29 dex. 
{The evolution of the black hole mass - stellar mass relation is assumed to follow a $(1+z)^2$  scaling  with redshift. Such an evolution is expected from theoretical arguments based on the black hole mass to halo circular velocity ($M_{\rm BH} - v_c$) relation: $M_{\rm BH} \propto v_c^4$  \citep[e.g.,][]{wyithe2002}, with $ v_{\rm c} \propto (1 + z)^{1/2}$, and assuming that the stellar mass to host halo mass relation does not evolve strongly with redshift \citep{behroozi2019}. The above scaling is borne out by recent observations \citep[e.g.,][]{caplar2015}. }

This allows to express the best-fitting relation \eq{lbolmbh}, and its associated scatter, in terms of the stellar mass and redshift of the host galaxy, with the scatter coming from the combination of that in the $L_{\rm Bol} - M_{\rm BH}$ and in the $M_{\rm BH} - M_*$ relations.

The expression for $L_{\rm peak}$ in \eq{ptdestart} is then given by \citep[e.g.,][]{fialkov2017}:
\begin{equation}
 L_{\rm peak}(M_*,z) = 133 \left(\frac{M_{\rm BH} (M_*, z)}{10^{6} M_{\odot}}\right)^{-1.5} L_{\rm Edd}  \, ,
\end{equation}
and the rate of TDEs, $\Gamma_{\rm TDE}$ is given by the observationally motivated fitting form {for core-dominated galaxies}
\citep[e.g.,][]{stone2016}:\footnote{We are making the implicit assumption that the TDE 
rate is influenced by physical processes that are not expected to change with 
redshift. The latest compilations connecting TDE rates to their observed host galaxy properties \citep[e.g.,][]{french2020} find that a TDE rate of $\sim 10^{-5}$  per year per galaxy is well borne out by recent data.}:
\begin{equation}
 \Gamma_{\rm TDE} (M_*, z) = 1.2 \times 10^{-5} \left(\frac{M_{\rm BH}(M_*, z)}{10^8 M_{\odot}}\right)^{-0.247} {\mathrm{yr}}^{-1} \, .
 \label{tderatepergalaxy}
\end{equation}
We also have
\begin{equation}
t_{\rm peak}(M_*, z, \epsilon) = 0.5 \epsilon M_{\odot} c^2/L_{\rm 
peak}(M_*,z) \, ,
\end{equation}
assuming a solar mass star is totally disrupted, consistently with Eq. (10) of 
\cite{fialkov2017}. Half the mass of the star is assumed to be accreted into the black hole which has a radiative efficiency $\epsilon$.

\begin{figure}
 \begin{center}
  \includegraphics[width = \columnwidth]{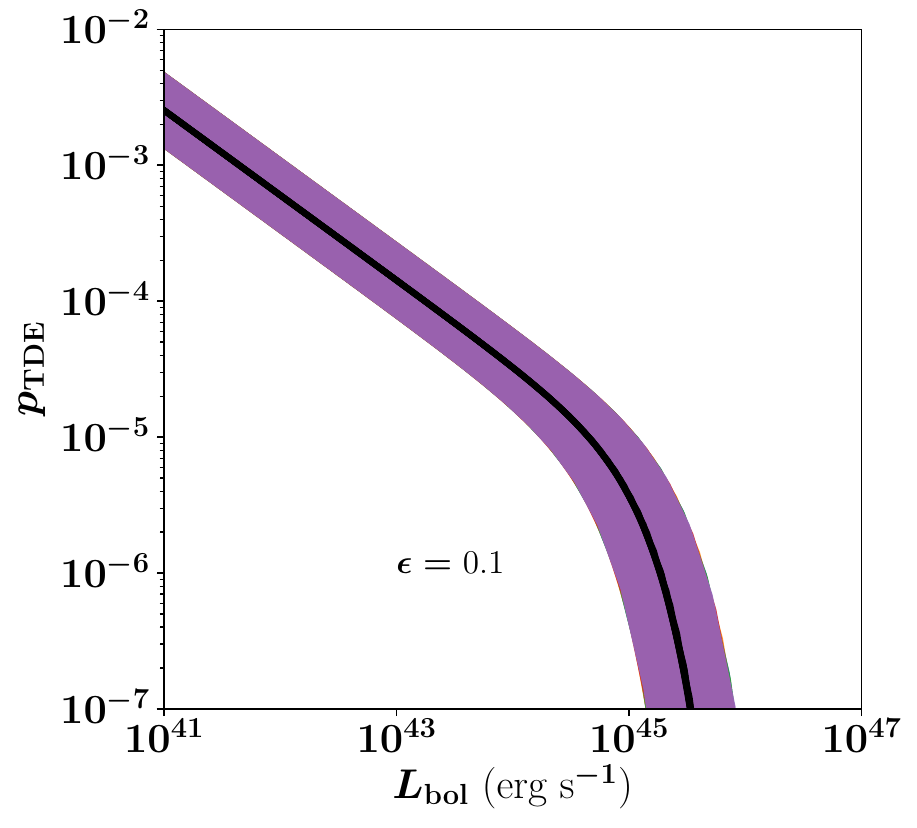}
 \end{center}
\caption{{ {The probability of TDE, $p_{\rm TDE}$ arising in AGN as a function of 
their bolometric luminosity, $L_{\rm bol}$ for a population having an assumed radiative efficiency $\epsilon = 0.1$.}}}
\label{fig:ptde}
\end{figure}

Using the above ingredients, \eq{ptdestart} can be recast as:
\begin{eqnarray}
 && p_{\rm TDE} (M_*, z;  \epsilon) = \Gamma_{\rm TDE} (M_*, z) 
t_{\rm peak}  (M_*, z, \epsilon) \nonumber \\
 &\times& \exp(-L_{\rm bol}  (M_*, z)/L_{\rm peak} (M_*, z))
\nonumber \\
 && \times \left(\frac{L_{\rm bol}   (M_*, z)}{L_{\rm peak}   (M_*, z)}\right)^{-1/\gamma_{\rm TDE}}/\gamma_{\rm TDE}  \, ,
\label{ptdemz}
\end{eqnarray}
with  $\epsilon$ being an adjustable parameter.
Fig. \ref{fig:ptde} plots the quantity $p_{\rm TDE} (M_*, z, \epsilon)$ against $L_{\rm bol} (M_*, z)$ along with the associated scatter, assuming $\epsilon = 0.1$ and for a range of host galaxy stellar masses between $\{10^{6}, 10^{15}\} M_{\odot}$. The figure shows that for a fixed $L_{\rm 
bol}$, the $p_{\rm TDE}$ is relatively insensitive to $z$ (since it 
has very little dependence on the black 
hole mass $M_{\rm BH}$, e.g. \citet{merloni2015}.) 
We also note the 
characteristic drop-off of this function above bolometric luminosities $\gtrsim 
10^{45}$ erg/s, which arises due to the exponential cutoff above $L_{\rm Bol} \sim L_{\rm peak}$ that corresponds to black hole masses greater than 
$M_{\rm BH} \sim 1.2 \times 10^8 M_{\odot}$\citep[e.g.,][]{kesden2012}.
The duration $t_{\rm peak}$ as a function of $L_{\rm bol}$ for these parameters and stellar masses ranges from about 0.1 to a few ten years. 

\begin{figure}
 \begin{center}
  \includegraphics[width = \columnwidth]{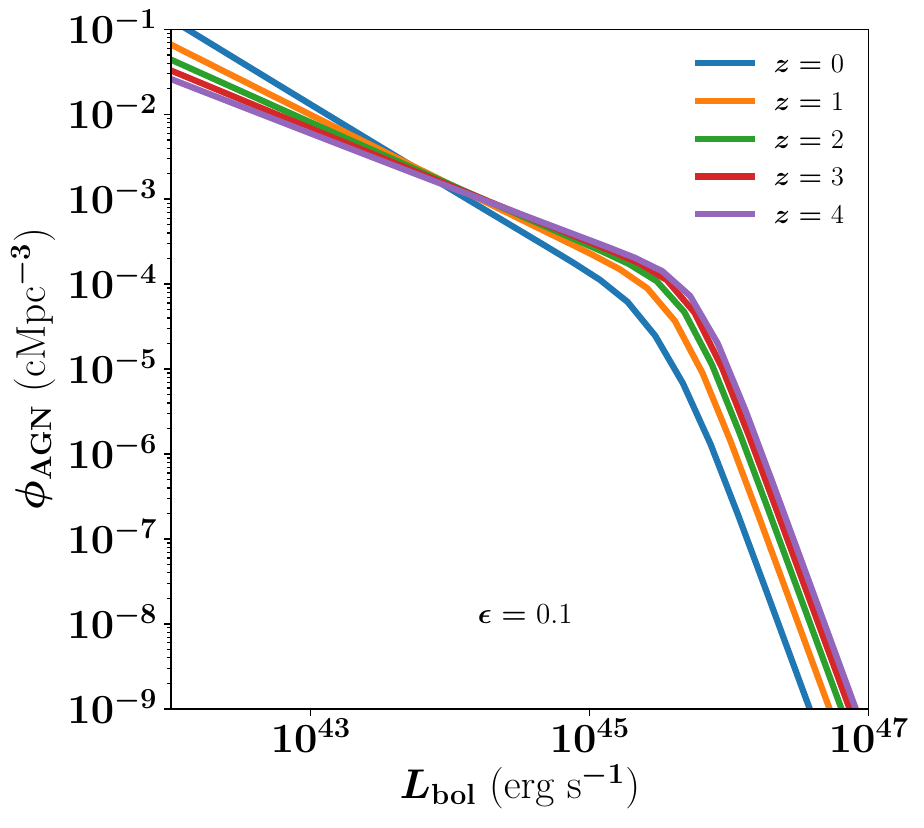} \includegraphics[width = \columnwidth]{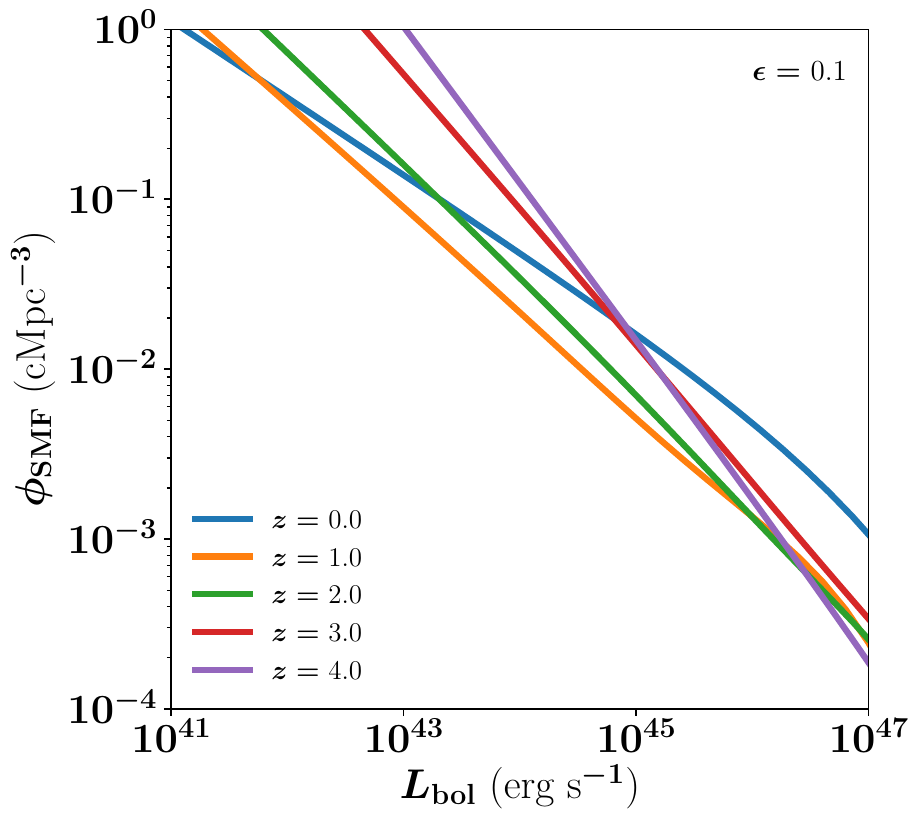}
 \end{center}
\caption{\textit{Top panel}: Observed AGN luminosity function, $\phi_{\rm AGN} (L_{\rm Bol}(M_*, z))$ at different redshifts from the X-ray data of \citet{aird2015} and using the bolometric correction of \citet{runnoe2012} to convert X-ray luminosity to bolometric luminosity. \textit{Lower panel}: Stellar 
mass functions compiled from the latest WINGS \citep[][$z \sim 0$]{vulcani2011} and COSMOS \citep[][$z \sim 0.2-5$]{davidzon2017} survey data converted into the equivalent $\phi_{\rm SMF}(L_{\rm Bol})$, by using the relation between $L_{\rm Bol}$ and $M_{*}$ via $M_{\rm BH}$.}
\label{fig:phiagnall}
\end{figure}

\section{Fraction of TDEs in AGN}
\label{sec:data}
Given the rates for TDEs occurring in AGN as calculated in the previous section, 
we can now compute the probability of TDEs occuring in different AGN populations 
in a particular bolometric luminosity interval as a function of redshift. To do 
this, we use the definition of $p_{\rm AGN}$, used in, e.g., 
\citet{merloni2015}, which measures the probability of AGN with a specified 
bolometric luminosity, $L_{\rm Bol}$, arising in host galaxies  with a given total 
stellar mass ($M_*$) at redshift $z$. This is given by  
\begin{equation}
 p_{\rm AGN} = \phi_{\rm AGN}(L_{\rm Bol} (M_*, z))/\phi_{\rm SMF}({M_*}|L_{\rm Bol}) \, ,
 \label{pagn}
\end{equation}

{ {The numerator $\phi_{\rm AGN}$ in the above expression denotes the AGN luminosity function, which is typically given by a 
Schechter or double power law form\footnote{The X-ray luminosity for 2 - 10 keV 
is converted to bolometric luminosity by using the bolometric correction $\log 
L_X  = (\log L_{\rm bol} - 23.04)/0.52$ from \citet{runnoe2012}.}(e.g. 
\citet{aird2015}, and shown in the top panel of Fig. \ref{fig:phiagnall}) {with the different colors indicating the various redshifts}.
  The denominator
$\phi_{\rm SMF}$ denotes the stellar mass function of the host galaxy, which is a function of the stellar mass and redshift.
Using the latest results from the WINGS \citep[at $z \sim 
0$;][]{vulcani2011} and the COSMOS \citep[at $z \sim 0.2 - 5$;][]{davidzon2017} 
for the stellar mass function, and the conversion between $L_{\rm bol}$ and $M_*(z)$ (through the $M_{\rm BH}-M_*$ relation, as outlined in Sec. \ref{sec:formalism}), we can express the $\phi_{\rm SMF}$ as a function of $L_{\rm bol}$. This is illustrated by
the curves in the lower panel of Fig. \ref{fig:phiagnall}.}} {The curves show that the `knee' of the stellar mass 
function gradually shifts to higher bolometric luminosities at higher redshifts, 
since a given stellar mass $M_*$ corresponds to a larger $M_{\rm BH}$ (and thus a larger $L_{\rm Bol}$) as $z$ 
increases.}

\begin{figure}
 \begin{center}
  \includegraphics[width = \columnwidth]{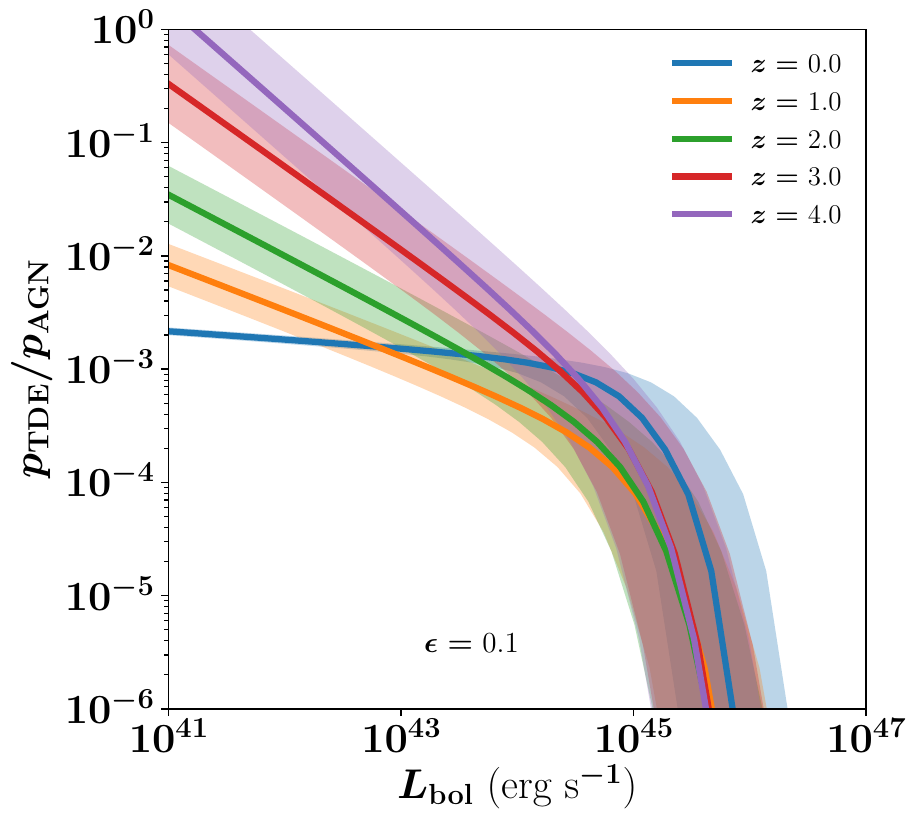}
 \end{center}
\caption{{ {Ratio $p_{\rm TDE}/p_{\rm AGN}$ for an assumed population of AGN having a radiative efficiency $\epsilon = 0.1$, at various redshifts. The shaded areas indicate the uncertainty induced by the scatter in the $L_{\rm bol} - M_*$ relation at various redshifts.}}}
\label{fig:pagnfid}
\end{figure}

Eqs. \ref{ptdemz} and \ref{pagn} can now be used to calculate the fraction of TDE {contributing to} AGNs as a function of their host halo and galaxy properties: 
\begin{equation}
 p({\rm TDE|AGN}) (M_*, z) = p_{\rm TDE} (M_*, z;  \epsilon)/p_{\rm AGN} (M_*,z) \, ,
 \label{probtdegivenagn}
\end{equation}
where the numerator on the right-hand side denotes the probability of TDEs occurring in AGN 
as calculated in the previous section (\eq{ptdemz}), or, equivalently, the joint probability of TDEs and AGN, and the denominator measures the 
probability of AGN occurring in host galaxies 
at different redshifts.\footnote{Note that this makes the implicit assumption that SMBH accrete with the same efficiency in the TDE and the non-TDE -- i.e. when fed by long-term gas accretion from the ISM  -- phase. While this may not be strictly true in general, it is the simplest possible scenario consistent with the available data, when taking into account that the exact value of the accretion efficiency does not change the results significantly.} For a fiducial AGN population having $\epsilon = 0.1$, the quantity $p({\rm TDE|AGN}) (M_*, z)$ is plotted against $L_{\rm Bol}(M_*,z)$ in Fig. \ref{fig:pagnfid}.

At low luminosities ($L_{\rm Bol} \sim 10^{43}$ erg/s), TDEs may account for a few percent of AGN at $z \sim 0-1$, consistent with observational estimates. However, as Fig. \ref{fig:pagnfid} shows, this number could reach a significant fraction of low-luminosity AGN at higher redshifts, which can be tested by future data. { {We see that TDEs could account for all the AGN seen in the $10^{41} - 10^{42}$ ergs/s luminosity regime at $z \gtrsim 3$, even with a significant factor `to spare'.}}

\section{TDEs in different AGN populations}
\label{sec:populations}
Thus far, we have considered the occurrences of TDEs in fiducial AGNs with specified host galaxy properties, which are assumed to be representative of a generic AGN population at various redshifts.  Next, we briefly consider the occurence of TDEs in a couple of specific classes of AGN, namely (i) Changing-Look AGN \citep[or Changing Look Quasars, CLQs; e.g.][]{yang2018, macleod2019, ross2018, stern2018} and (ii) Compton-Thick AGN or CT AGN \citep[e.g.][]{Balokovi2014, malizia2009, marchesi2018}.

The origin of the so-called `changing-look' phenomena in AGN, in which a luminous quasar dims significantly over a timescale of $< 10$ years, is currently unknown.
AGN classified as changing-look (CL) have been currently discovered in the local universe up to a highest redshift of $z \sim 2$. 
More than 
20\%
\citep{macleod2019} of quasars with $L_{\rm bol} > 10^{44}$ erg s$^{-1}$,
and 30-50\% of quasars with $L_{\rm bol} \sim 10^{45} - 10^{47}$ erg s$^{-1}$
\citep{rumbaugh2018} are found to fall into this category.
{ {While several CLQs have Eddington ratios around 0.1 (e.g., Fig. 14 of \citet{Graham2019}), in the present study we assume that the Eddington ratio distribution of the CLQs does not differ from that of the broader AGN population.  

We can thus estimate how many of the low-$z$ CLQ discovered so far may be accounted for as TDE {flares contributing to}
AGN. 
From Fig. \ref{fig:pagnfid}, we see that at $z \sim 1$, a maximum of $\sim 1\%$ of the AGN with $L_{\rm bol} \gtrsim 10^{42}$ erg  s$^{-1}$
may originate from TDE-flares, which thus represents up to about $5 \%$ of the observed
CLQ abundance.\footnote{The
$t_{\rm peak}$ timescales for the $L_{\rm bol}$ values considered here range from about 0.1
to a few ten years over $z \sim 0-4$, which
 are broadly consistent with the changing-look timescales expected from the literature.}}}

\begin{figure}
 \begin{center}
  \includegraphics[width = \columnwidth]{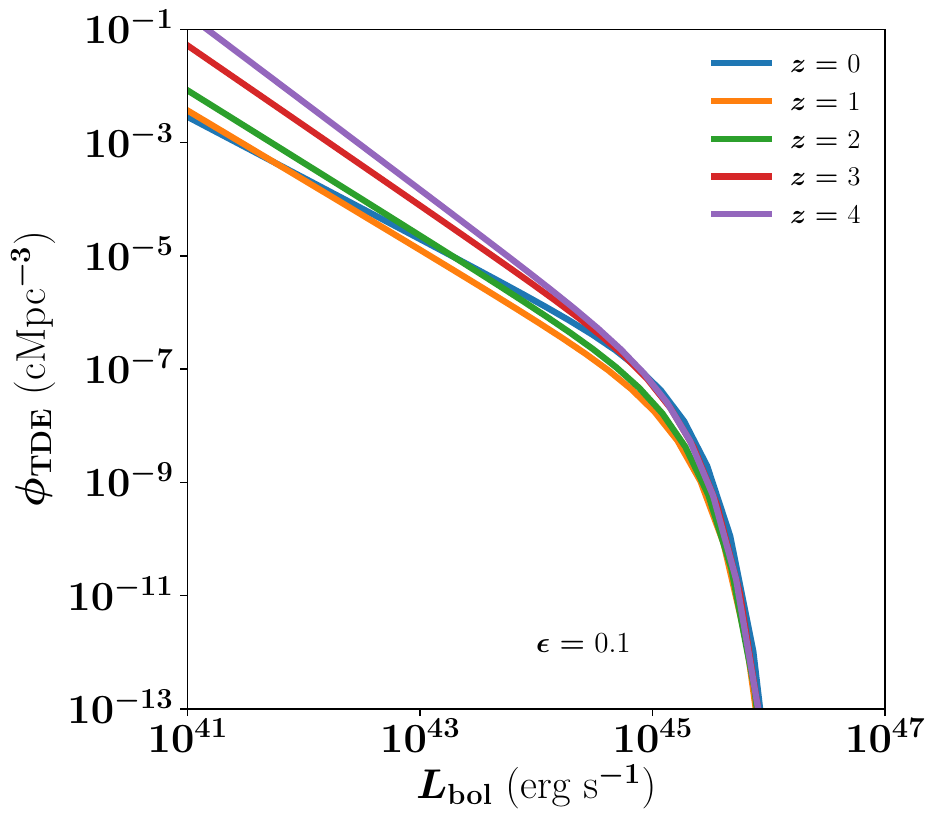}
 \end{center}
\caption{{  {Predicted luminosity function of TDE {flares contributing to} AGN, $\phi_{\rm TDE} \equiv p_{\rm TDE} \times \phi_{\rm SMF}$ for an assumed population having radiative efficiency $\epsilon = 0.1$. The predicted $\phi_{\rm TDE}$ serves as a lower limit on the luminosity function of changing look quasars, assuming their intrinsic parameters remain redshift independent.}}}
\label{fig:phitde}
\end{figure}

On the other hand, at higher redshifts, there is evidence to indicate a much larger fraction of TDE flares contributing to AGN in the CLQ population.
The first CLQs at $z > 2$ which have recently been discovered \citep{ross2019} have Eddington ratios of $\eta > 0.05$. Upcoming experiments such as the Dark Energy Spectroscopic Instrument \citep[DESI;][]{desi2016} and LSST will enable stringent constraints on this population at higher redshifts. Assuming the intrinsic parameters of the population do not change significantly with redshift, we predict a `TDE luminosity function', $\phi_{\rm TDE} \equiv p_{\rm TDE} \times \phi_{\rm SMF}$ as shown in Fig. \ref{fig:phitde}.
This function provides a lower limit on the abundance of CLQs at high-$z$ {whose flares may be exclusively contributed by TDEs}.

\begin{figure}
 \begin{center}
  \includegraphics[width = \columnwidth]{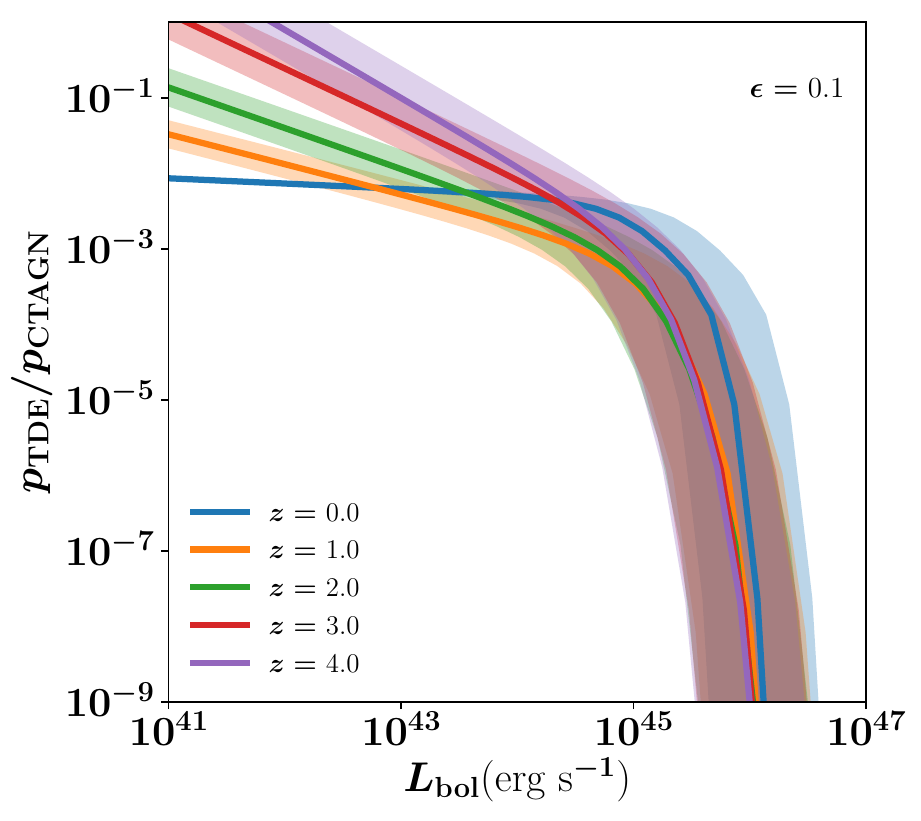}
 \end{center}
\caption{{ {The probability of CT AGN hosting a TDE at various redshifts. The 
assumed AGN population has a radiative efficiency of $\epsilon = 0.1$. The shaded areas indicate the uncertainty induced by the scatter in the $L_{\rm bol} - M_*$ relation at various redshifts.}}}
\label{fig:ctagn}
\end{figure}

{ {The  discussion of TDEs is of particular interest to Compton Thick (CT) environments because the TDE debris in tidal streams has a large column density and is naturally Compton-thick. This population may thus appear as CT AGN; with the thick envelope of gas making them detectable in X-rays but not in optical-UV.}} Adopting the luminosity function of CT AGN (as a function of the absorbed AGN fraction) from \citet{aird2015}, we calculate the ratio $p_{\rm TDE}/p_{\rm CTAGN}$ which is plotted in 
Fig. \ref{fig:ctagn}. We find that the TDE fraction in CT AGN can be a {  {few percent}} 
around $L_{\rm bol} \sim 10^{43}$ erg  s$^{-1}$ at redshifts $z \sim 1$. At higher redshifts, it may reach a { {greater (up to several ten percent)}} fraction of the CT AGN population.  In all 
cases, it drops
sharply as expected around $L_{\rm bol} \sim 10^{46}$ erg  s$^{-1}$.

\section{Discussion}
\label{sec:conclusions}
We have considered the fraction of AGN which may arise from tidal disruption events (TDEs) at low and high redshifts. { {In so doing, we have extended the locally observed relations connecting the rates of TDEs to the BH host masses of the AGN through their bolometric luminosities \citep{nobuta2012,kormendy2013}, to high redshifts using the inferred evolution of the black hole mass {to host galaxy stellar mass relation} \citep{caplar2015, wyithe2002}. Combining these findings with the observationally derived galaxy luminosity function \citep{vulcani2011, davidzon2017}, we have estimated the fraction of TDEs occurring in different AGN populations. For a fiducial AGN population having an average radiative efficiency of $\epsilon = 0.1$, we find that TDEs are able to account for a few percent of all AGN with bolometric luminosities $L_{\rm Bol} \gtrsim 10^{42}$ erg  s$^{-1}$ at $z \sim 1$, but may reach several tens of percent at $z \gtrsim 3$.}}

It is also of interest to explore to what extent specific AGN populations are likely to be the result of TDE flares at low and high redshifts. Two examples of such populations are the so-called Changing-Look AGN (or Changing Look Quasars, CLQs) and Compton-Thick AGN (or CT AGN). On applying the above formalism to these populations, we found that a maximum of about a few percent of the observed CLQ AGN (having $L_{\rm Bol} > 10^{42}$ erg  s$^{-1}$) may be consistent with a TDE {contribution to their origin} at $z \sim 1$, though this fraction may increase rapidly at higher redshifts, { {to several tens of percent}} at $z \gtrsim 3$. We also find that in the TDE fraction may be about a few percent of CT AGN with $L_{\rm Bol} \gtrsim 10^{43}$ erg  s$^{-1}$ at redshifts 
$z < 1$, but may again account for { {up to several}} ten percent of this population at $z \gtrsim 3$.

{In Fig. 10 of \citet{lanzuisi2018}, an analysis of CT AGN carried out from the \textit{Chandra}-COSMOS legacy survey finds that the fraction of CT AGN increases with increasing redshifts, and the merger fraction in CT AGN increases with increasing luminosities and redshifts. Interestingly, a sharp rise in the ``fraction of merged/disturbed morphology" in 
these AGN occurs above $L_{\rm Bol} > 10^{46}$ erg  s$^{-1}$. This may be attributable to the  non-TDE triggered AGN being triggered primarily by mergers, with the merger fraction increasing sharply once the maximum black hole mass of $10^{8} M_{\odot}$ is reached, consistently with our results in Fig. \ref{fig:ctagn}.}

{ {Throughout the analysis, we have considered various sources of scatter in the in the adopted relations. The average trend between the bolometric luminosity and the black hole mass contains an associated scatter, which is estimated directly from the observations \citep{nobuta2012}. This scatter is added to that in the relation connecting the black hole mass and host galaxy stellar mass \citep{kormendy2013}. A source of scatter we have not taken into account is that coming from the bolometric correction, which may have an uncertainty of  about 15-20\% \citep{runnoe2012}. However, we note that this uncertainty is chiefly driven by our knowledge of the bolometric correction, rather than an intrinsic scatter due to physical processes.}}

A caveat to the analysis here  is that it is not presently clear whether the black hole-galaxy correlations are valid for less massive galaxies (with $M_{\rm BH}$ below $10^9 M_{\odot}$)  at $z \sim 0$ \citep[e.g.,][]{shirakata2017, nguyen2020} due to the lack of understanding of seed black holes. The occupation fraction of black holes in less massive galaxies, if lower than the average \citep[e.g.,][]{volonteri2008, volonteri2010, greene2012} would influence the TDE detection rate and therefore the TDE fraction in AGN. The black hole-galaxy stellar mass connections may also have an evolutionary dependence weaker than the currently employed $(1+z)^2$, though this remains  a point of active debate.

{{The present analysis assumes that the intrinsic per-galaxy TDE rate (given by \eq{tderatepergalaxy}) remains roughly constant across redshifts. While observations of TDEs have found rates of $10^{-5}$ TDEs per year per galaxy in samples out to redshift $z \sim 0.4$ \citep{french2020}, this is still an order of magnitude smaller than that predicted by theoretical models \citep[e.g.,][]{kochanek2016, stone2016}. 
 The above findings also illustrate that the detection of TDEs in high-redshift AGN  can place constraints on the black hole mass function and its evolution assumed by the theoretical models \citep[e.g.,][]{hopkins2007}, including the consequences of mergers \citep{shankar2009}.
}}

Our results serve as a useful benchmark for calibrating future simulations of the physical processes giving rise to TDEs in AGN at high redshifts, and their inferred contributions in various AGN populations. Current and future surveys, e.g. the \textit{Athena}\footnote{https://www.the-athena-x-ray-observatory.eu/} and \textit{Lynx}\footnote{https://www.lynxobservatory.com/} missions that target different AGN populations will be able to place stringent constraints on the CT AGN fractions at high redshifts, and combining this data with the findings of spectroscopic surveys such as the LSST and DESI will allow us to better quantify the contributions of TDEs as AGN at these epochs. The above formalism will  in turn enable the most precise constraints thus far on the evolution of the central black hole - halo mass relation of the host galaxies at high redshifts. Further, the forthcoming detections of TDEs will unveil the as-yet poorly understood processes governing the growth of IMBHs and SMBHs, whose mergers are expected to constitute the primary gravitational wave event triggers detectable by the forthcoming LISA observatory.

\section{Acknowledgements}
 HP acknowledges support by the Swiss National Science Foundation (SNSF) Ambizione Grant PZ00P2\_179934 ``\textit{Probing the Universe: through reionization and beyond'}'. The work of AL was supported in part by the Black Hole Initiative at Harvard University, which is funded by grants from the JTF and GBMF.

\def\aj{AJ}                   
\def\araa{ARA\&A}             
\def\apj{ApJ}                 
\def\apjl{ApJ}                
\def\apjs{ApJS}               
\def\ao{Appl.Optics}          
\def\apss{Ap\&SS}             
\def\aap{A\&A}                
\def\aapr{A\&A~Rev.}          
\def\aaps{A\&AS}              
\def\azh{AZh}                 
\def\baas{BAAS}
\def\jcap{JCAP}
\def\jrasc{JRASC}             
\def\memras{MmRAS}
\def\na{New Astronomy}
\def\nat{Nature}
\def\mnras{MNRAS}             
\def\pra{Phys.Rev.A}          
\def\prb{Phys.Rev.B}          
\def\prc{Phys.Rev.C}          
\def\prd{Phys.Rev.D}          
\def\prl{Phys.Rev.Lett}       
\def\pasp{PASP}               
\def\pasj{PASJ}
\def\physrep{Phys. Repts.}
\def\qjras{QJRAS}             
\def\skytel{S\&T}             
\def\solphys{Solar~Phys.}     
\def\sovast{Soviet~Ast.}      
\def\ssr{Space~Sci.Rev.}      
\def\zap{ZAp}                 
\let\astap=\aap
\let\apjlett=\apjl
\let\apjsupp=\apjs

% for the bibliography, at the end 
\bibliographystyle{aa} % style aa.bst 
\bibliography{mybib} % your references Yourfile.bib 

\label{lastpage}

\end{document}